\documentstyle[aps,prl,twocolumn,epsf,epsfig]{revtex}
\def\beq{\begin{equation}}
\def\eeq{\end{equation}}
\def\beqarr{\begin{eqnarray}}
\def\eeqarr{\end{eqnarray}}

\begin{document}
\draft
\twocolumn[\hsize\textwidth\columnwidth\hsize\csname @twocolumnfalse\endcsname

\title{A disordered RKKY lattice mean field theory for ferromagnetism in 
diluted magnetic semiconductors}
\author{D. J. Priour, Jr, E. H. Hwang, and S. Das Sarma}
\address{Condensed Matter Theory Center, Department of Physics,
University of Maryland, College Park, MD 20742-4111}


\maketitle

\begin{abstract}
We develop a lattice mean field theory for ferromagnetic 
ordering in diluted magnetic semiconductors by taking into account the 
spatial fluctuations associated with random disorder in the magnetic 
impurity locations and the finite mean free path associated with low 
carrier mobilities.  Assuming a carrier-mediated indirect  
RKKY exchange interaction among the magnetic impurities,  
we find substantial deviation from the extensively used
continuum Zener model Weiss mean-field predictions.  Our theory allows accurate
analytic predictions for $T_{c}$, and provides simple 
explanations for a number of observed anomalies 
including the non-Brillouin function magnetization curves, the suppressed 
low-temperature magnetization saturation, and the dependence of $T_c$
on conductivity.
\end{abstract}

]

Much of our current understanding of ferromagnetism in diluted 
magnetic semiconductors (DMS), most notably in the extensively studied
molecular beam epitaxy grown Mn-doped $\textrm{Ga}_{1-x}\textrm{Mn}_{x}
\textrm{As}$ (with the Mn 
doping level $x \approx 0.01-0.1$ system~\cite{uno}), has
been based on a very simple continuum Weiss mean field theory 
approximation~\cite{dos}  
of the Zener model for the local (p-d) exchange coupling between 
the impurity magnetic moment $S = 5/2$ $d$-levels of Mn, and the itinerant
carrier spin polarization, $s = 3/2$ holes of $p$-character in the valence 
band of GaAs.  Spatial fluctuations associated with the random locations of 
Mn local moments  
are neglected in this continuum Weiss mean field theory (MFT). 
There is no theoretical rationale, except 
simplicity, underlying the neglect of the strong quenched spatial 
disorder due to the random magnetic impurity locations.  Indeed, there
have been several Monte Carlo simulations~\cite{tres,cuatro,cinco,seis,siete}
attempting to include
spatial disorder effects in the theory.  These 
simulations as well as a recently developed percolation 
theory~\cite{ocho,nueve}
have explicitly demonstrated the manifest importance of quenched disorder
in DMS ferromagnetism, at least for the localized insulating DMS systems.
There are also strong experimental signatures for the important
interplay between disorder and magnetism in DMS materials.

In this Letter, we develop 
the first systematic theory for DMS ferromagnetism explicitly including 
spatial disorder effects in the successful Zener-RKKY mean field 
model [2,9].  Our
theory is explicitly constructed for the metallic DMS systems with 
itinerant carriers (since we assume the carrier-mediated effective 
Mn-Mn indirect magnetic exchange interaction to be of free carrier 
mediated RKKY 
form), but the inclusion of a finite carrier mean free path in the theory
(as we do in addition to the inclusion of Mn spatial disorder) allows us
to make specific predictions about the dependence of the magnetic behavior
of the system on the carrier transport properties.   
We also include a direct nearest-neighbor Mn-Mn  
antiferromagnetic exchange interaction in our disordered RKKY-mean field 
theory, which takes on significance for larger Mn concentrations.  An 
associated essential feature of our model, which may be of crucial 
importance, is the fully discrete nature of our disordered mean field
theory on a lattice. 
It is well-known that the discreteness
associated with the specific lattice structure introduces unique 
features to the RKKY exchange interaction which are not caught in the 
corresponding continuum approximation.  Our model, although conceptually 
simple, is actually quite rich as it depends on five independent physical
parameters of the DMS system:  the carrier-induced Mn-Mn RKKY coupling
($J_{0}$), the direct Mn-Mn (nearest-neighbor) antiferromagnetic exchange
coupling ($J^{AF}$), the local moment density ($n_{i}$), the free
carrier density ($n_{c}$), and the carrier mean free path ($r_{0}$).
In principle, the randomness in the Mn locations on the lattice could
also be parameterized,  
particularly if clustering (or other 
spatial correlations) of Mn impurities is important during GaMnAs 
growth.  We neglect at this stage any such correlation in Mn 
spatial locations (since no independent experimental information on 
the nature of quenched disorder is available), assuming the Mn atoms 
to be uniformly randomly distributed in the zinc-blende GaAs lattice 
at the Ga substitutional sites. (It would be easy to incorporate a 
more detailed description of disorder in the theory and to include Mn 
defects such as interstitials.)  We also assume that the five 
parameters (i.e. $J_{0},J^{AF},n_{i},n_{c},r_{0}$) of our theory are
completely independent of each other, which is a theoretically and 
physically consistent approximation within our model since these 
parameters must be determined experimentally (or otherwise) and are 
independent parameters for our model.

Our effective Hamiltonian
describes the Mn-Mn magnetic interaction between classical Heisenberg
spins ${\bf S}_{i}$ on a lattice:
\begin{equation}
{\mathcal H} = \sum_{ij} J^{RKKY}_{ij} {\bf S}_{i} \cdot {\bf S}_{j} + 
\sum_{ij}J^{AF}_{ij} {\bf S}_{i} \cdot {\bf S}_{j},
\label{Eq:eq1}
\end{equation}
where ${\bf S}_{i}$ is the $i$th Mn local moment of spin $5/2$; $J^{AF}_{ij}$
is the direct antiferromagnetic exchange interaction between 
nearest-neighbor Mn spins, i.e. $J^{AF}_{ij} = 0$, unless $i$ and $j$ are
nearest neighbors.  The first term in the effective Hamiltonian, the 
carrier-mediated RKKY indirect exchange interaction responsible for 
producing DMS ferromagnetism, describes the magnetic interaction between 
the Mn local moments induced by the extended (free) carrier spin 
polarization (these free carriers happen to be the valence band or the 
impurity band holes in GaMnAs).  This indirect Mn-Mn exchange interaction
arises from~\cite{uno,dos,nueve} the local Zener coupling (or the so-called p-d
hybridization) between the holes and the Mn $d$-levels, which then leads 
to the effective Mn-Mn RKKY interaction:
\begin{equation}
J^{RKKY}_{ij}
\equiv J_{0} r^{-4}\left[\sin(2k_{F}r) - 2k_{F}r\cos(2k_{F}r)\right]
\label{Eq:eq2}
\end{equation}
where $r = \left| {\bf R}_{i} - {\bf R}_{j} \right|$ is the spatial 
separation between the magnetically coupled Mn atoms; $k_{F} \propto
n_{c}^{1/3}$ is the Fermi wavevector corresponding to the carrier 
density $n_{c}$, and the RKKY coupling strength $J_{0}(>0)$, taken 
as a parameter in our theory, is related to the local Zener coupling 
$J_{pd}$ between the Mn local moments and the hole spins, $J_{0} \propto
m J_{pd}^{2}$, where
$m$ is the hole effective mass.  Note 
that for low carrier densities the RKKY exchange interaction is mostly
ferromagnetic except for large Mn-Mn separation ($r \geq k_{F}^{-1}$),
and therefore, as long as $n_{c} \ll n_{i}$, where $n_{i}$ is the active 
density of Mn local moments (i.e. typical $r \propto n_{i}^{-1/3}$), the 
frustration effects associated with the oscillatory nature of RKKY exchange
interaction are unimportant in the problem; distinguishing the DMS 
systems from  random dilute
metallic magnetic alloy spin glass systems (e.g. Cu-Mn), which are
typically in the  opposite limit of $n_{c} \gg n_{i}$.
It is important 
to emphasize that a continuum Weiss mean-field theory (which necessarily 
neglects all spatial fluctuation effects by averaging over the Mn 
positions), blindly applied to the Mn-Mn RKKY interaction, as was already 
done~\cite{diez} a long time ago (and revived recently [2] 
in the context of DMS 
systems), always yields long-range ferromagnetic ordering of the Mn local 
moments for all values of $n_{c}$ and $n_{i}$, with a mean-field 
ferromagnetic transition temperature 
$T_{c}^{MFT} \propto J_{0}n_{i}n_{c}^{1/3}$.
This is obviously incorrect for larger values of $n_{c}$ where ferromagnetism 
would eventually disappear~\cite{once} in a lattice model.  

     We include the collisional broadening due to the finite carrier mean 
free path ($r_{0}$) in the theory by incorporating an exponential suppression 
$e^{-r/r_{0}}$ in Eq.~(\ref{Eq:eq2}), indicating that the RKKY 
interaction is cut 
off for $r \gg r_{0}$ due to resistive scattering effects. (This is
important in view of the strong experimentally observed correlation
between $T_c$ and conductivity typically found in DMS systems.) This form for 
the collisional broadening induced suppression of the RKKY interaction is 
theoretically well-justified~\cite{doce} for studying ferromagnetic ordering 
of the local moments.  We 
have also considered thermal effects in the RKKY interaction by 
generalizing Eq.~(\ref{Eq:eq2}) to the corresponding finite 
temperature formula,
but typically $T_{c} \ll T_{F}$ for GaMnAs, where $T_{F}$ ($T_{c}$) is 
the carrier Fermi
temperature (Curie temperature), and therefore thermal broadening of the RKKY 
interaction is quantitatively and qualitatively unimportant in our
calculations  
in contrast to the collisional broadening effects included in the
$e^{-r/r_{0}}$ 
suppression.  Note that an important novel feature of our theory is the 
``integrating-out'' of the free carrier variables as we consider an effective 
(disordered) local moment magnetic Hamiltonian defined by Eq.~(\ref{Eq:eq1}), with 
the free carrier information entering the theory only through the various
physical parameters of the model, e.g. $J_{0},n_{c},r_{0},m$, etc.
This simplification is well-justified in studying DMS ferromagnetism since the 
long-range ferromagnetic ordering in the system arises entirely from the Mn local
moments with the itinerant holes contributing little to ferromagnetism (by virtue
of $n_{i} \gg n_{c}$ and $S_{Mn} > s_{hole}$).

     The Hamiltonian of Eq.~(\ref{Eq:eq1}) can be rewritten as a generalized 
Heisenberg model for Mn (classical) spins on a disordered GaAs lattice 
${\mathcal H} = \sum_{ij}J_{ij}(r){\bf S}_{i}\cdot{\bf S}_{j}$ (where
$J_{ij} \equiv J^{RKKY}_{ij} + J^{AF}_{ij} \equiv J(r)$) with the sum over $i$,$j$
extending over (random) magnetic impurity locations in the fcc GaAs lattice.  With
no loss of generality we approximate the short-ranged antiferromagnetic interaction
$J_{ij}^{AF}$ by subsuming it in the definition of $J_{ij}^{RKKY}$ or $J(r)$ by 
multiplying it by a factor of $1/2$ if the impurities occupy nearest-neighbor
positions in the fcc lattice.  This avoids introducing an additional unknown
parameter $J^{AF}$, which could easily be introduced if such a need arises.

    We use a lattice mean field theory and consider each impurity spin 
    $\bbox{S}_{i}$ to be immersed in an effective magnetic field,
$B_{\rm{eff}}^{(i)} = \frac{1}{g\mu_{B}}\left( J_{i} \langle \langle
S^{z}\rangle \rangle \right)$, where $J_{i} \equiv \sum_{j} J_{ij}$ is
the sum of all couplings to the impurities surrounding the site $i$, and 
$\langle \langle S^{z} \rangle \rangle$ is the thermally and site-averaged
polarization, and $g$ is the g-factor corresponding to the impurity. 
Clearly, $J_{i}$ depends on the impurity configuration and $B_{\rm{eff}}^{(i)}$
is specific to the position $R_{i}$.  Given $B_{\rm{eff}}^{(i)}$, the thermally    
averaged spin polarization $\langle S^{z}_{i} \rangle$ is ($k_{B} = 1$)
$\langle S_{i}^{z} \rangle = SB_{S}\left(g\mu_{B}B_{\rm{eff}}^{(i)}/T \right)$,
where $B_{S}$ is the usual mean field thermal Brillouin function.
Although we calculate
in mean field and use the average spin $\langle \langle S^{z} \rangle \rangle$,
 we retain the site dependence in 
$J_{i}$, thereby taking into account impurity disorder which subjects different 
Mn spins to different couplings depending on the local impurity configurations as 
determined by the random Mn locations on the GaAs lattice.
$\langle S_{i}^{z} \rangle$ is numerically 
calculated for a specific impurity distribution.  To determine  
$\langle \langle S^{z} \rangle \rangle$, the polarization averaged over
all impurities, we integrate 
over all possible realizations of disorder, obtaining 
\begin{equation}
\langle \langle S^{z} \rangle \rangle = \int P(J) SB_{S}\left(\langle \langle S^{z} 
\rangle \rangle J/T \right) dJ,
\label{Eq:eq5}
\end{equation}
where $P(J)$ is the probability distribution of $J$.  
One then calculates $\langle \langle S^{z} \rangle \rangle$ self-consistently
from Eq.~(\ref{Eq:eq5}).
P(J) is determined a priori via Monte Carlo sampling.  
We assume that the Mn impurities occupy Ga lattice sites with uniform probability
as determined by its concentration $x$ 
in $\textrm{Ga}_{1-x}\textrm{Mn}_{x}\textrm{As}$.

In our lattice mean field theory the ferromagnetic transition temperature 
$T_{c}$ is determined by the site-averaged exchange coupling and is given by
\begin{equation}
T_{c} = \frac{35}{12} x \sum_{i=1}^{\infty} N_{i} J(r_{i}),  
\label{Eq:eq6}
\end{equation}
where $N_{i}$ and $r_{i}$ are the numbers and distances of the $i$th nearest
neighbors, respectively.  We express $r_{i}$ and the remaining two length scales 
$r_{0},(2k_{F})^{-1}$ in units of the fcc unit cell length.  The continuum
limit is reached
for $r_{0},(2k_{F})^{-1} \ll 1$ with
\begin{eqnarray}
&T_{c}^{cont}& = \left( 140\pi x/3 \right) 
\int_{0}^{\infty} r^{2} J(r) dr\\
&=&T_{c}^{MFT}\left[1 - \tan^{-1}(2k_{F}r_{0})/(2k_{F}r_{0}) \right],
\label{Eq:eq7}
\end{eqnarray}
where $T_{c}^{MFT} \equiv (280\pi J_{0}x/3)(3\pi^{2}n_{c}/2)^{1/3} \propto
J_{0}xn_{c}^{1/3}$ is the continuum Weiss MFT 
value for $T_{c}$ which has been employed~\cite{dos}  extensively in the recent 
DMS literature (note that $S = 5/2$ here).  
The factor
$f(2k_{F}r_{0}) \equiv [1-\tan^{-1}(2k_{F}r_{0})/(2k_{F}r_{0})]$ takes into account 
the exponential cutoff imposed
by the finite mean free path.   
We recover the continuum Weiss MFT result 
when $2k_{F}r_{0} \gg 1$, i.e. in the strongly metallic regime.  However, when 
$r_{0}$ becomes comparable to the length scale 
$\left(2k_{F} \right)^{-1}$; $f(x) \approx \frac{1}{3}x^{2} - \frac{1}{5}x^{4}
\cdots$, the RKKY interaction is effectively suppressed
and $T_{c}$ is substantially lower than 
$T_{c}^{MFT}$ even in the continuum approximation.  
To obtain an accurate formula for $T_{c}$, it is necessary also to take 
into account the 
antiferromagnetic interaction between Mn impurities and to correct
for the differences between the continuum approximation of Eq.~(\ref{Eq:eq7}) and 
the discrete lattice sum of Eq.~(\ref{Eq:eq6}).  In the same way that an integral
and a discrete approximation to that integral differ by a power series
in the step size, 
the difference between the continuum and discrete formulae described 
above can be written as a series in $k_{F}$.  With these improvements, 
one finds as a reasonable large $r_{0}$ approximation for $T_{c}$ 
\begin{eqnarray}
\nonumber
T_{c} &=& T_{c}^{MFT}(1+\alpha_{2}n_{c}^{2/3} + \alpha_{4}
n_{c}^{4/3} + \alpha_{6}n_{c}^{2} + \cdots)\\ 
\nonumber
&\times&(1-\beta_{1}r_{0}^{-1}[n_{c}^{1/3}(1-\beta_{2}r_{0}^{-1})]+r_{0}^{-1}[ 
\beta_{3} - \beta_{4}r_{0}^{-1}]) \\ 
\label{Eq:eq8}
&\times& f(2k_{F}r_{0})- \frac{35}{2}xJ(1/\sqrt{2});\\
\alpha_{2} &=& -.733256, \alpha_{4} = 6.2594 \times 10^{-2},\nonumber\\
\alpha_{6} &=& 2.89 \times 10^{-3}, \alpha_{8} = 3.7 \times 10^{-4}; \nonumber\\
\beta_{1} &=& .1633,\beta_{2} = .4284, \beta_{3} = .5584, \beta_{4} = .1176 
\end{eqnarray}
$T_{c}$ in Eq.~(\ref{Eq:eq8})
is correct to better than one part in $10^6$ for large $r_{0}$ (as compared with 
our numerical calculations).  Accuracy
diminishes as $r_{0}$ is made smaller, though the formula given in 
Eq.~(\ref{Eq:eq8}) is still correct to within $1\%$ for 
$r_{0}=1$. 
One can also arrive at a formula to cover the small 
$r_{0}$ regime.  In this case, $T_{c}$ can be expressed as the sum of 
two terms 
\begin{eqnarray}
T_{c} &=& \frac{35}{12k_{B}} x\left[6J(1/\sqrt{2}) + 16\pi \int^{\infty}_{r_{a}}J(r)r^{2}dr \right], 
\label{Eq:eq9}
\end{eqnarray}
where the first term in brackets is the exact nearest neighbor contribution while
the second term contains the remaining couplings calculated in the 
continuum limit, and the optimal choice for $r_{a}$ is $0.953$.  
For $r_{0} \leq 1$ the relative error for $T_{c}$ from 
Eq.~(\ref{Eq:eq9}) is less than $2\%$.
Evidently, the 
first term (accounting for the effect of impurities on neighboring 
sites) dominates as $r_{0}$ becomes comparable to the size of the 
GaAs unit cell.  In this regime, the 
antiferromagnetic coupling removes half of $T_{c}$ (within our 
simplistic choice for $J^{AF}$) and thus plays
a very important role for small mean free path.  Obviously the 
result here depends strongly on the specific approximation one
uses for $J^{AF}$ and would be stronger for stronger 
Mn-Mn AF coupling.

\begin{figure}
\centerline{\epsfig{figure=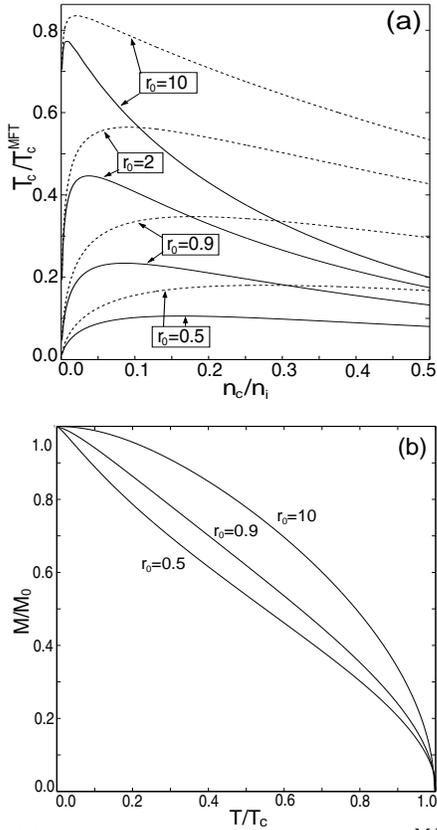,width=2.25in}}
\caption{(a) $T_{c}$ relative to continuum Weiss $T_{c}^{MFT}$ for
carrier mean free path $r_{0}=10$, $2$, $0.9$, and $0.5$
Solid curves
are plotted with the antiferromagnetic interaction taken into
account, while dashed curves depict $T_{c}$ with antiferromagnetism
suppressed.  (b) Magnetization profiles corresponding to
$r_0=10$ ($x=0.05$), $r_0=0.9$ ($x=0.01$), $r_0=0.5$ ($x=0.015$);
$n_{c}/n_{i} = 0.3$ for all $r_{0}$ values}
\label{Fig:fig-2}
\end{figure}

In Fig.~\ref{Fig:fig-2}(a) we    
show our direct numerical calculation of $T_{c}$
as a function of the carrier density $n_{c}$ (for fixed
Mn doping level $x$) for four different values of the carrier mean free
path varying from strongly metallic ($r_{0} = 10$) to ``almost''
insulating ($r_{0} = .5$).  Clearly, for most choices 
of $n_{c}$ and $r_{0}$, $T_{c}^{MFT}$ is a poor approximation for $T_{c}$ in 
the disordered lattice system.
We emphasize that the strong dependence of our calculated $T_c$ on the
system conductivity (through $r_0$) is consistent with GaMnAs
experimental results where increasing conductivity is found to enhance $T_c$.
In Fig~\ref{Fig:fig-2}(b) we show our calculated spontaneous magnetization $M(T)$
results, which depend on the full exchange distribution $P(J)$.
The $M(T)$ profile (convex, concave, or linear) depends on whether 
the system is in the insulating (small $r_{0}$, small $n_{c}$) or 
metallic (large $r_{0}$, large $n_{c}$) regime.  Concavity in $M(T)$ 
is a signature of an insulating system, 
while convex profiles appear deep in the metallic regime [9].  
For intermediate impurity densities and mean free paths, it is 
possible to obtain a linear magnetization curve.  Within the 
framework of our model, we are able to span both the localized 
and metallic regimes by having very small and large values of $r_{0}$,
respectively, and we find both types of behavior
in the magnetization profiles as can be seen in Fig.~\ref{Fig:fig-2}(b).

There is a strong correlation
between the degree of concavity in $M(T)$ 
and the extent to which the coupling probability distribution $P(J)$ has 
a strong weight near zero interaction strength and a multimodal 
profile.  The multimodal $P(J)$ in general leads to a concave $M(T)$.
A reasonable measure of the 
concavity is $\gamma \equiv \int^{t2}_{t1}M^{''}(T)dT$, or the difference in 
the slopes of $M(T)$.  The sign of $\gamma$ indicates
whether $M(T)$ is convex (negative $\gamma$), 
concave (positive $\gamma$), or linear
(if $\gamma \approx 0$).  The temperatures $t_{1}$ 
and $t_{2}$ are chosen to capture an intermediate temperature range,
neither very close to $T_{c}$ nor to zero.  To go beyond 
the influence of the low temperature behavior, we set

\begin{figure}
\centerline{\epsfig{figure=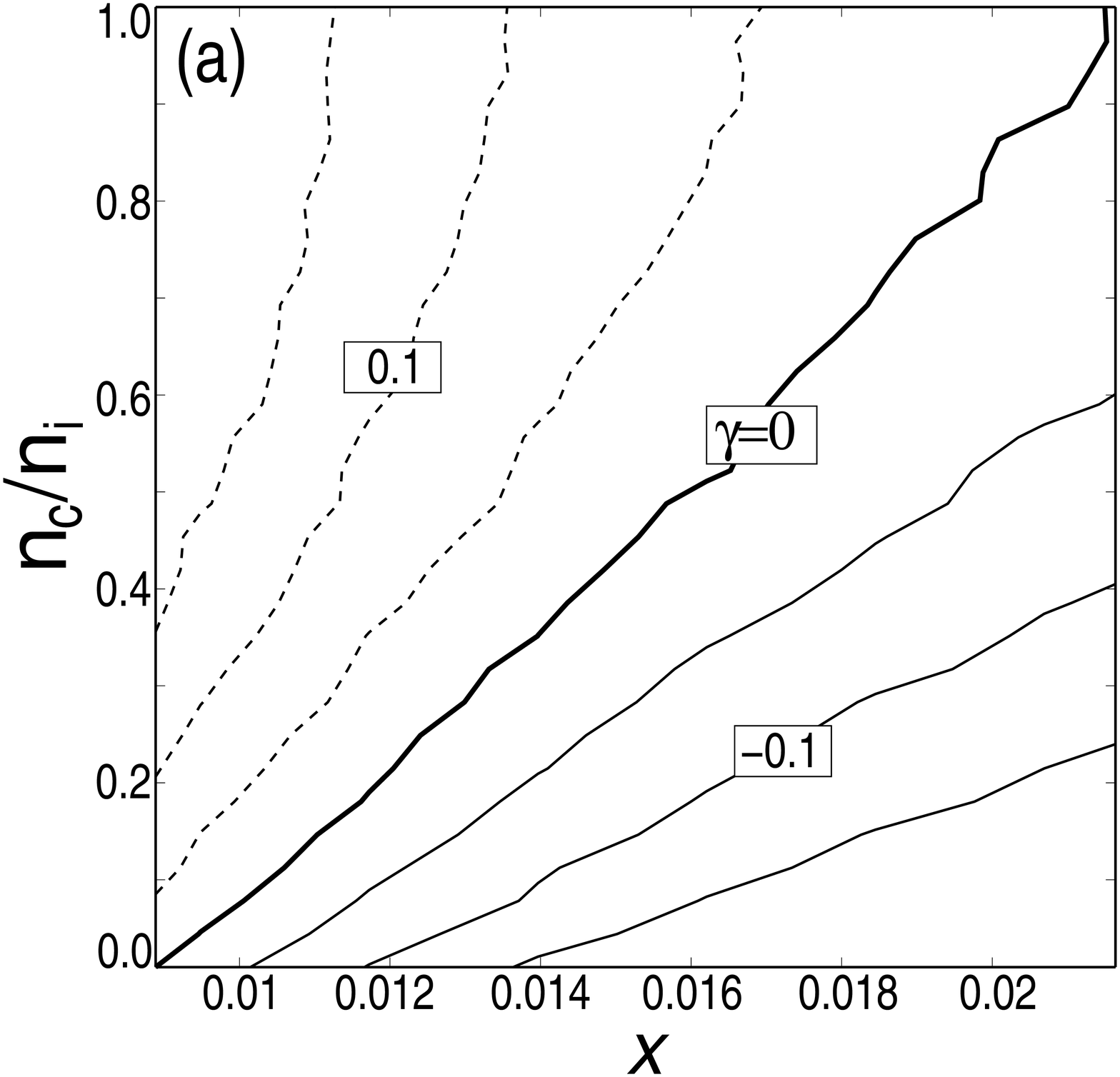,width=2.2in}}
\centerline{\epsfig{figure=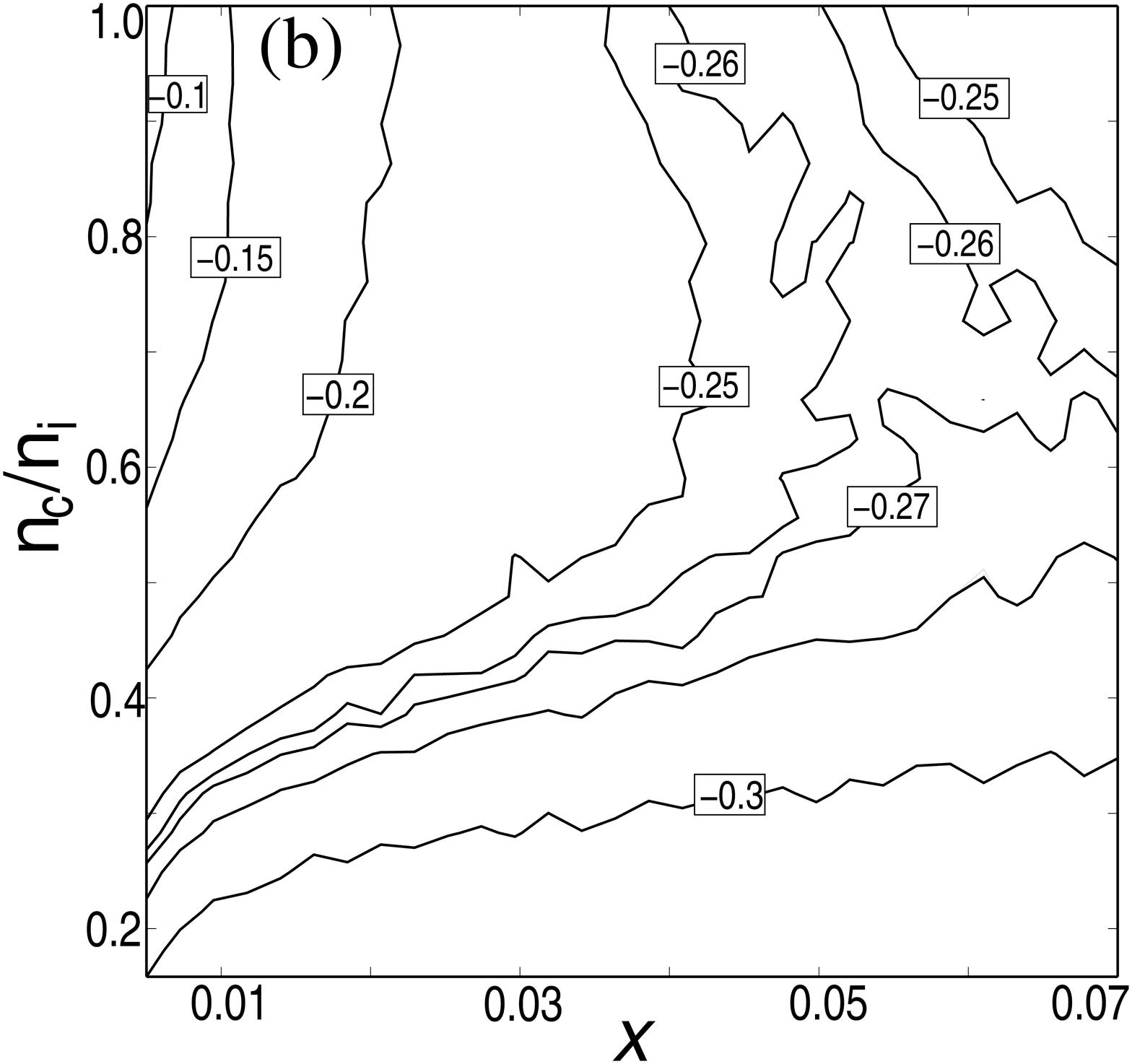,width=2.2in}}
\caption{Magnetization phase diagram: contour plots of the concavity
  parameter $\gamma$ as a function of 
Mn concentration $x$ and $n_{c}/n_{i}$ for two values of carrier mean free
path (a) $r_{0} = .5$ and (b) $r_{0} = 2$, where $n_{c}$ ($n_{i})$ are
the carrier 
(local moment) densities, and $x$ is the Mn doping level.  The $\gamma
= 0$ contour divides the concave/convex 
phases in (a).}
\label{Fig:fig2}
\end{figure}

\noindent
$t_{1} = .1t_{c}$; to avoid the critical region, we use $t_{2} = .7t_{c}$.
This result is a rough criterion for concavity, since the sign of $\gamma$ 
indicates whether the profile is convex, concave, or linear.  
In Fig.~\ref{Fig:fig2} we give results for our ``magnetization phase
diagram'' where the regimes of convex/concave magnetization behavior
are depicted on the $n_c/n_i$--$x$ two dimensional plots.
Contour plots in Fig.~\ref{Fig:fig2} show in a concise way important 
trends; one sees very clearly the transition from some concavity  in $M(T)$ to 
convex $M(T)$ behavior with increasing $r_{0}$.
Raising impurity density also tends to 
make the magnetization profile more convex.   

We have also calculated the saturation magnetization $M_{0} \equiv
M(T\rightarrow0)$ 
using our theory, finding that consistent with experimental
observations $M_{0}$  
could indeed be less than unity particularly for larger values of
relative carrier density ($n_{c}/n_{i}$) and/or for more metallic systems (i.e.
larger values of $r_{0}$).  
The main suppression mechanisms are direct AF 
coupling between nearest-neighbor Mn-Mn interaction (increasing with
$J^{AF}$ and  $x$) and the 
oscillatory nature of the RKKY exchange coupling at large values 
of $k_{F}r$ (increasing with $n_{c}/n_{i}$).

To summarize, we have developed a disordered lattice mean field theory 
for DMS ferromagnetism which 
incorporates spatial fluctuations associated with random lattice 
locations of the impurity moments,
the finite carrier mean free path, and the Mn-Mn nearest-neighbor
antiferromagnetic coupling.
We calculate $T_{c}$ and 
magnetization curves for $\textrm{Ga}_{1-x}\textrm{Mn}_{x}\textrm{As}$
ferromagnetic semiconductors, finding that all magnetic 
properties,
including $T_{c}$, depart significantly from the predictions of the  
continuum Weiss mean-field theory.
A particularly salient feature of our results is the strong
theoretical correlation between $T_c$ and the metallicity of the
system (i.e. $r_0$) as observed experimentally.
It should be feasible to incorporate 
spin-orbit coupling and detailed valence band structure in the theory
by generalizing our simple single valence band RKKY model.  The most
important essential approximation of our model is the assumption of 
the RKKY form for the carrier-mediated indirect exchange interaction 
between the impurity local moments.  Recent first 
principles band structure calculations~\cite{trece} show that the effective DMS
interaction between impurity moments is indeed of the RKKY form.  
There have also been recent numerical calculations explicitly 
establishing~\cite{catorce} the validity of RKKY coupling in
disordered semiconductors.  
Our theory is computationally relatively efficient allowing us to
obtain DMS magnetic properties as a function of five independent
physical parameters ($J_0$, $J^{AF}$, $n_c$, $n_i$, $r_0$). We believe
that the lattice mean field approximation is fairly sound by
virtue of $n_i \gg n_c$ in the DMS systems (and because of the long
range nature of the ferromagnetic RKKY interaction).

\end{document}